\documentclass[%
 aip,
 amsmath,amssymb,
 reprint,%
]{revtex4-1}
\usepackage{hyperref}
\usepackage{graphicx}% Include figure files
\usepackage{dcolumn}% Align table columns on decimal point
\usepackage{bm}% bold math
\usepackage[utf8]{inputenc}
\usepackage[T1]{fontenc}
\usepackage{mathptmx}
\usepackage{etoolbox}

\makeatletter
\def\@email#1#2{%
 \endgroup
 \patchcmd{\titleblock@produce}
  {\frontmatter@RRAPformat}
  {\frontmatter@RRAPformat{\produce@RRAP{*#1\href{mailto:#2}{#2}}}\frontmatter@RRAPformat}
  {}{}
}%
\makeatother
\begin{document}

\preprint{AIP/123-QED}

\title[Sample title]{Backscattering-Immune Floquet Conversion in Ring Modulators}
% Force line breaks with \\
\author{Awanish Pandey}
 \email{opcawanish@iitd.ac.in}

 \altaffiliation{Optics and Photonics Centre, Indian Institute of Technology, New Delhi}%Lines break automatically or can be forced with \\
 \author{Alex Krasnok}
\altaffiliation{Department of Electrical and Computer Engineering, Florida International University, Miami, Florida 33174, USA}

\date{\today}% It is always \today, today,
             %  but any date may be explicitly specified

\begin{abstract}
Backscattering in micro-ring cavities induces mode mixing and limits device performance. Existing methods to mitigate backscattering often involve complex fabrication processes or are insufficient for complete suppression. In this work, we introduce a novel method to eliminate backscattering by operating the cavity at an exceptional point (EP). By engineering non-conservative coupling between degenerate clockwise (CW) and counter-clockwise (CCW) modes, we achieve chiral transmission that prevents degeneracy lifting and suppresses unwanted mode coupling. Unlike previous approaches that rely on precise gain-loss balance or complex structures, our method utilizes non-conservative coupling between the counterpropgating cavity modes. Using this method, we further show significant enhancement in the cavity performance in Floquet mode conversion efficiency at the EP. Our highly adaptable approach enables seamless integration into various photonic platforms with electro-optic modulators. This advancement mitigates backscattering and improves the precision of light-matter interactions, offering promising applications in quantum communication and information processing.
\end{abstract}

\maketitle

Optical microring cavities are fundamental components in photonics, widely employed in applications ranging from classical and quantum information processing to optical communication and sensing~\cite{nh_intro1,nh_intro3,intro1,intro2}. A significant challenge in these devices is backscattering caused by imperfections on the resonator's surface or within its material. These imperfections lead to coupling between the inherently degenerate clockwise (CW) and counter-clockwise (CCW) propagating modes, resulting in light scattering into the counter-propagating mode~\cite{pandey_ol}. This unintended coupling can severely limit device performance. For instance, in laser gyroscopes operating at low rotational speeds, backscattering can induce undesired injection locking, reducing sensitivity~\cite{gyro}. In devices relying on nonlinear amplification or the interference of counter-propagating fields—such as sensors, optical computing elements, and isolators, backscattering can diminish efficiency and introduce reflections that degrade functionality~\cite{haye}. Various techniques have been proposed to mitigate backscattering, including fabrication improvements to reduce imperfections and using unidirectional elements~\cite{Kutsaev2021}. However, these methods can be complex or insufficient for complete suppression.

Exceptional points (EPs) offer a promising alternative for addressing backscattering issues. EPs are unique spectral singularities in non-Hermitian systems where both eigenvalues and their corresponding eigenvectors coalesce~\cite{ep_intro, Krasnok2021}. In photonics, EPs have been extensively studied in systems such as metamaterials, photonic crystals, and microcavities, revealing novel phenomena like unidirectional invisibility and enhanced sensitivity \cite{jlt_sensors,sym_brk}. Integrated photonics provides an ideal platform for exploring EPs due to the ability to precisely engineer material absorption, radiation leakage, and mode interactions through device design. This has facilitated the development of several miniaturized optical structures operating at an EP such as a silicon microring resonators with Mie scatterers \cite{ep_silicon}, and the infinity-loop micro-resonator \cite{ep_infinity}.

\begin{figure*}[htbp]
\centering
 \includegraphics[width=150 mm]{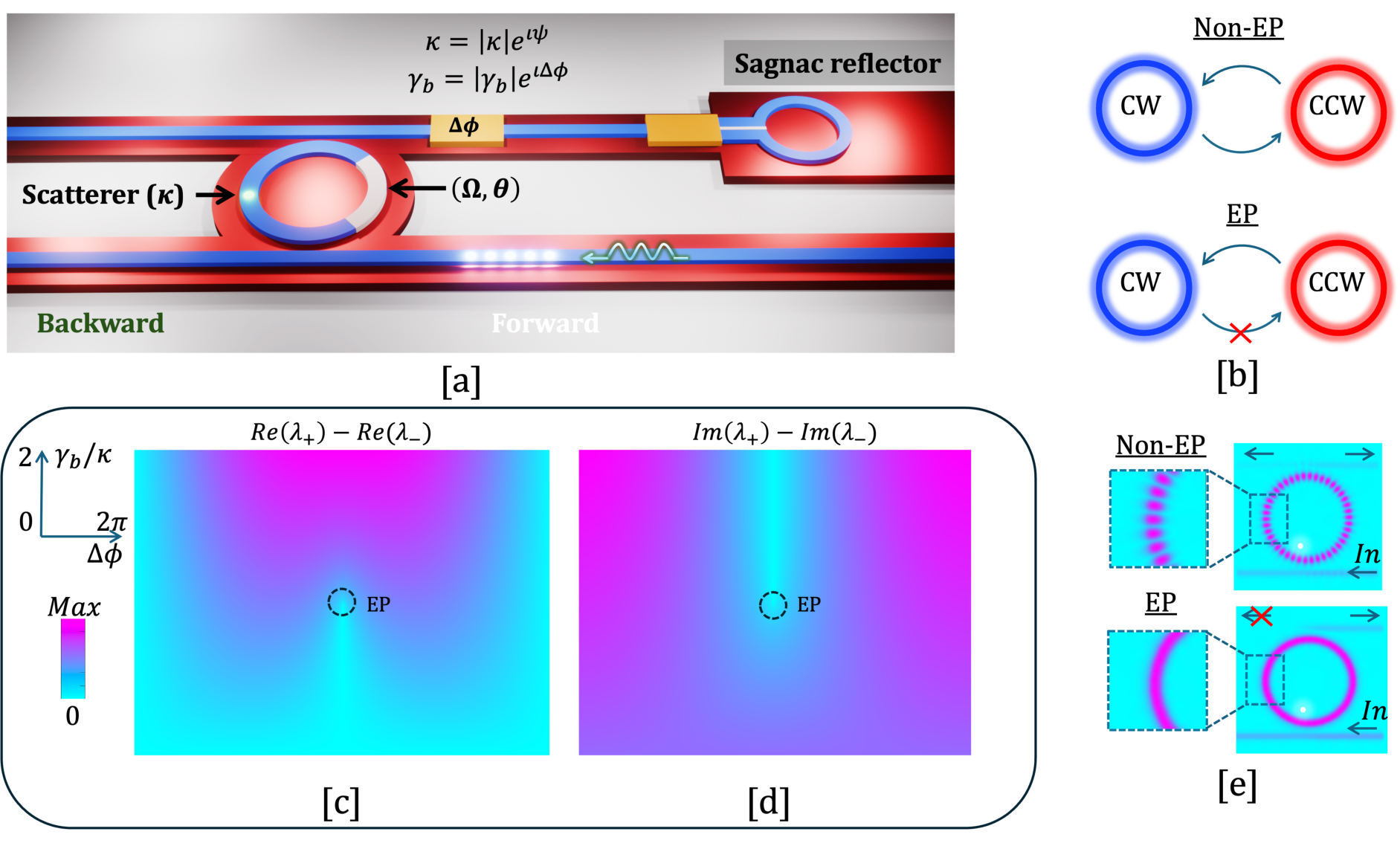}
 \caption{[a] Schematic of the proposed structure, [b] coupling between CW and CCW at a non-EP and on EP, [c] and [d] variation of difference between real and imaginary part of the eigen-frequencies once the degeneracy is lifted, and [e] resonance E-field distribution at a non-EP and on EP clearly showing a traveling wave type distribution at EP and a standing wave type distribution at a non-EP.}\label{fig:schematic}
\end{figure*}

In this paper, we present a novel approach to suppress backscattering in optical microring cavities by operating at an exceptional point induced through non-conservative coupling between the counter-propagating modes. Unlike previous methods, our approach does not require gain elements or precise gain-loss balancing. Instead, we employ a combination of techniques, including the excitation of the counterclockwise (CCW) mode within the microring cavity through internal back-reflection, as well as external back-reflection enabled by a Sagnac reflector. Using this technique, we demonstrate a highly efficient Floquet mode conversion~\cite{fleury} mechanism by subjecting the cavity to a carefully designed modulation scheme that exploits the EP's characteristics and optimizes key operating parameters of the cavity. This method leverages standard electro-optic modulators and is compatible with existing integrated photonics platforms, making it practical for experimental implementation. Our analysis includes detailed theoretical modeling and numerical simulations that quantitatively demonstrate the suppression of backscattering and the efficiency of Floquet mode conversion.

Figure~\ref{fig:schematic}(a) illustrates the schematic of the examined structure. It comprises a ring cavity that can inherently support two degenerate resonant modes traveling in clockwise (CW) and counterclockwise (CCW) directions with the same eigenfrequency ($\omega_0$) but orthogonal eigenvectors \cite{bogaerts}. The intrinsic loss inside the cavity is denoted by $\gamma_l$, and the coupling strength with the bus waveguides is $\gamma_c$. The total loss is given by $\gamma = \gamma_l + \gamma_c$, and the resonance mode field amplitude inside the cavity is represented by $\alpha(t)$. 

To model the lifting of degeneracy, we consider backscattering induced by imperfections, which is typically distributed along the cavity but can be approximated as a lumped scatterer~\cite{lumped}. This backscattering introduces a coupling $\kappa$ between the CW and CCW modes. In practice, a non-zero $\kappa$ is obtained either by changing the topology of the cavity or by adding Rayleigh backscattering~\cite{haye} in the form of a defect in the cavity waveguide. The back-scattering element has a coupling strength of $\kappa$ with a phase $\phi$.

We also introduce another mechanism for lifting degeneracy referred to as Autler-Townes splitting~\cite{autler_townes}, achieved by integrating a bi-directional back-scattering element inside the ring and a Sagnac-reflector in one of the two bus-waveguides, preceded by an embedded phase shifter ($\Delta\phi$) to realize a complex coefficient ($\gamma_b$), resulting in a non-conservative coupling between the CW and CCW modes i.e. the coupling coefficient is represented as a complex quantity, defined by both its coupling amplitude and phase. The Sagnac reflector, combined with the phase shifter can lead to a controlled non-reciprocal phase shift between the CW and CCW modes (see Fig. \ref{fig:schematic}(b)), effectively implementing the non-conservative coupling term $\gamma_b$ in our model.

The time-modulation is achieved by applying a microwave driving signal to the cavity~\cite{acs_pandey} that changes the resonance frequency according to $\omega_0(t) = \delta\omega_m\cos(\Omega t)$. Similarly, the loss is modulated as $\gamma_l(t) = \delta\gamma_l\cos(\Omega t + \theta)$~\cite{modulation}. In practice, such a modulation scheme can be realized by cascading a phase modulator with an amplitude modulator~\cite{modulation2}. The input optical signal is represented by $s_{in}(t)=Pe^{\iota\omega_l t}$, where $P$ is the source amplitude and $\omega_l$ is the angular frequency. The coupled mode equations describing the evolution of CW and CCW mode-field amplitudes $\alpha_{cw}(t)$ and $\alpha_{ccw}(t)$ are given by:
\begin{equation}
\frac{d\alpha_{cw}}{dt} = [\iota\omega_0(t) - \gamma(t)]\alpha_{cw} + \iota\kappa\alpha_{ccw} - \iota\sqrt{2\gamma_c}Pe^{\iota\omega_l t}
\label{eqn:cmt}
\end{equation}
\begin{equation}
\frac{d\alpha_{ccw}}{dt} = [\iota\omega_0(t) - \gamma(t)]\alpha_{ccw} + \iota(\kappa+\gamma_b)\alpha_{cw}
\label{eqn:cmt2}
\end{equation}

In the absence of modulation ($\delta\omega_m = 0$, $\delta\gamma_l = 0$), i.e., in a passive cavity, the eigenvalues are given by equations:
\begin{equation}
\lambda_{\pm} = \omega_0 \pm\sqrt{\kappa(\kappa + \gamma_b e^{\iota\Delta\phi})},\label{eqn:eigen}
\end{equation}
demonstrating the lifting of degeneracy between the CW and CCW modes, resulting in two distinct modes with different resonance frequencies and losses i.e. Re($\lambda_{\pm}$) and Im($\lambda_{\pm}$) respectively. Under the conditions $|\kappa| = |\gamma_{b}|$ and $\Delta\phi = \pi + \psi$, both the eigenvalues and eigenvectors of the system coalesce, forming an EP as depicted in Fig. \ref{fig:schematic}(c) and (d).  Figure~\ref{fig:schematic}(c) illustrates the difference between the real parts of the eigenvalues, representing the resonance frequency splitting, while Fig.~\ref{fig:schematic}(d) shows the difference in their imaginary parts, indicating the loss splitting, both of which become zero at the EP. 

At this EP, equation (\ref{eqn:cmt}) shows a non-zero coupling between the CCW mode and the CW mode ($\kappa$), but no coupling between the CW mode and the CCW mode ($\kappa + \gamma_b = 0$), resulting in chiral coupling. In effect, this indicates that once the counterclockwise (CCW) mode is excited, it fully transfers the optical power to the clockwise (CW) mode. However, a similar transfer of optical power from the CW mode to the CCW mode is inhibited, resulting in the coalescence of eigenvalues i.e. disapperance of the splitting i.e., the vanishing of the mode splitting. Fig.~\ref{fig:schematic}(e) shows the resonance mode field pattern in the cavity obtained via FDTD simulation. In the top part of Fig.~\ref{fig:schematic}(e), a standing wave type of field pattern is observed, which results from the bi-directional interaction between the CW and CCW modes under non-EP conditions. However, due to the non-reciprocal coupling between CW and CCW modes at the EP, a traveling wave type of mode field is obtained, confirming the suppression of backscattered waves \cite{ep_field}.

\begin{figure}[!t]
\centering
 \includegraphics[width=90 mm]{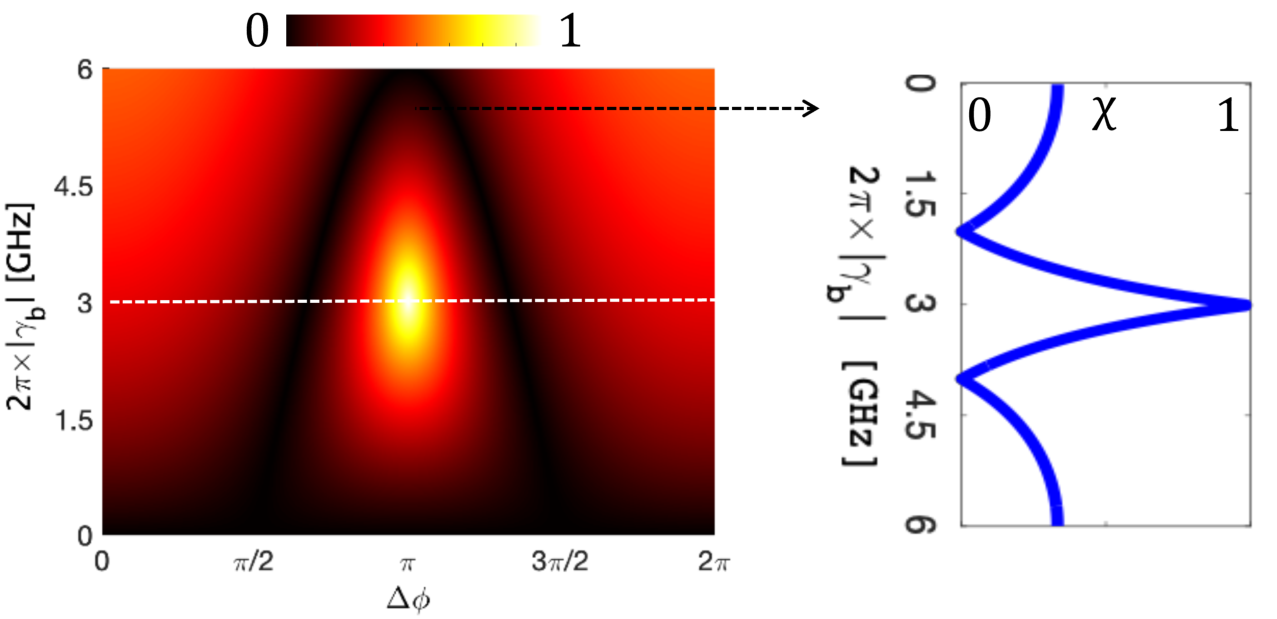}
 \caption{Variation of $\chi$ as a function of $\gamma_b$ and $\Delta\phi$. The chirality peaks when $|\gamma_b| = |\kappa|$ and the phase difference between them is $\pi$.}\label{fig:chiral}
\end{figure}

The asymmetrical coupling between the two counter-propagating modes is quantified by introducing a chirality parameter ($\chi$), which characterizes the reflections when the input signal is incident from opposite directions. When the CW (CCW) mode is excited, the resulting reflection is equal to $2\gamma_c|\alpha_{ccw(cw)}|^2$. By solving equations (\ref{eqn:cmt}) and (\ref{eqn:cmt2}) for a passive cavity, we obtain $\chi$ as the ratio of the difference to the sum of the output when the input signal is applied to excite the $CCW$ and the $CW$ mode, respectively. This yields the equation: 
\begin{equation}
 \chi = \frac{|\kappa|-\sqrt{|\kappa|^2+|\gamma_b|^2+2|\kappa||\gamma_b|\cos(\Delta\phi-\psi)}}{|\kappa|+\sqrt{|\kappa|^2+|\gamma_b|^2+2|\kappa||\gamma_b|\cos(\Delta\phi-\psi)}},
 \label{eqn:chirality}
\end{equation}
which explicitly relates $\chi$ to the system's coupling parameters. From equation (\ref{eqn:chirality}), it is clear that the terms within the square root symbol represent the magnitude squared of the sum of two complex numbers. This sum would be 0 if both complex numbers have the same magnitude and a phase difference of $\pi$ between them. For a completely symmetric coupling (e.g., $\gamma_b = 0$, $\kappa \neq 0$), the value of $\chi$ is zero as shown in Fig. \ref{fig:chiral}. However, $\chi$ assumes a non-zero value for asymmetric coupling, indicating the presence of chirality. At the EP, where the coupling between the CW and CCW modes is entirely suppressed, $\chi$ reaches unity, signifying maximal chirality and effective backscattering suppression.

\begin{figure}[!t]
\centering
 \includegraphics[width=90 mm]{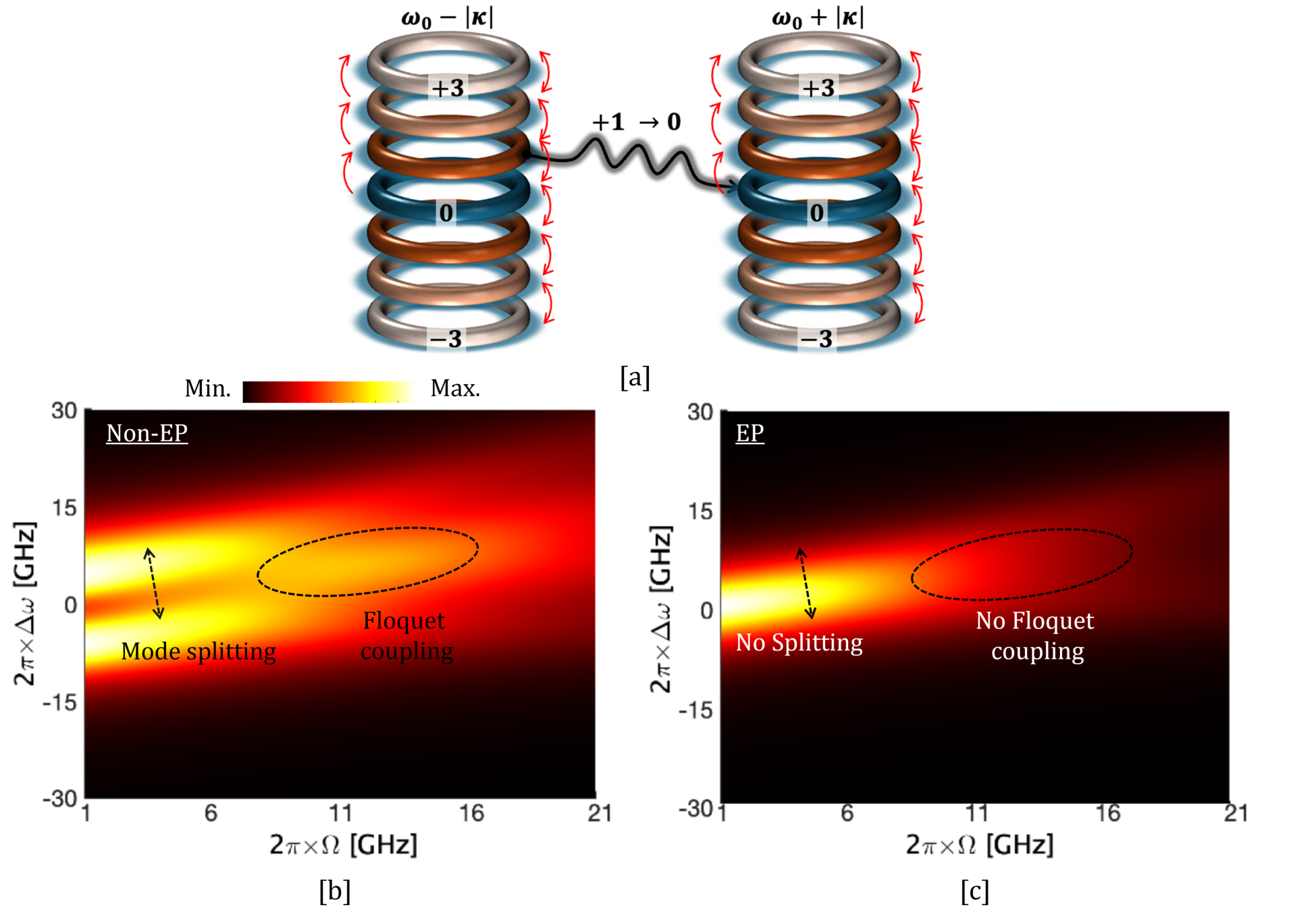}
 \caption{[a] Coupling between Floquet mode generated by the split resonances as well as the coupling between the first order Floquet mode from $\omega_0 - \kappa$ resonance with the $\omega_0 + |\kappa|$ resonance, [b] and [c] represent the variation of the first higher-order Floquet mode under non-EP and at EP.}\label{fig:sideband_contour}
\end{figure}

Subsequently, we examine the transmission characteristics of the cavity when a time-varying complex signal is applied to a section of the cavity, changing $\omega_0(t)$ and $\gamma_l(t)$ periodically with an angular frequency of $\Omega$. By employing the Floquet theorem in conjunction with coupled mode theory, the modal field amplitude of the $CW (CCW)$ mode can be expressed as a Fourier series: $\alpha_{cw(ccw)}(t) = \sum_{n=-\infty}^{\infty}\alpha_{cw(ccw)}^ne^{\iota\omega_l t + \iota n\Omega t}$ where $\alpha_{cw(ccw)}^n$ represents the time-independent complex amplitude of the $n^{th}$ Floquet harmonic from the CW and CCW modes. Substituting this expansion into the coupled mode equations, the temporal evolution of the CW and CCW modes is described by equations:
\begin{equation}
 [\iota\Delta\omega+\iota n\Omega+\gamma]\alpha_{cw}^n - \iota S^+\alpha_{cw}^{n-1}-\iota S^-\alpha_{cw}^{n+1} -\iota\kappa\alpha_{ccw}^n=-\iota\sqrt{2\gamma_c}P\delta_{n0},
 \label{eqn:recursive1}
\end{equation}
\begin{equation}
 [\iota\Delta\omega+\iota n\Omega+\gamma]\alpha_{ccw}^n - \iota S^+\alpha_{ccw}^{n-1}-\iota S^-\alpha_{ccw}^{n+1} -\iota(\kappa+\gamma_b)\alpha_{cw}^n=0,
 \label{eqn:recursive2}
\end{equation}
where $S^{\pm} = (\delta\omega_m + \iota\delta\gamma_l e^{\pm\iota\theta})/2$. From equations (\ref{eqn:recursive1}) and (\ref{eqn:recursive2}), it can be determined that, depending on the cavity and modulation parameters, the cavity can operate in three distinct regimes. The first regime occurs when $(\delta\gamma_b = 0, \Delta\phi=0)$, $\delta\omega_m \neq \delta\gamma_l$, and $\theta=0$. In this scenario, $|S^-| = |S^+|$ and the coefficients of $n\pm1$ are identical. As a result, the higher and lower order Floquet modes are symmetrically coupled, and the optical energy is spread out evenly between them, as shown by the red arrows in Fig.~\ref{fig:sideband_contour}a. A notable feature in this regime is when $\Omega$ becomes comparable to the separation between the cavity split resonance (2$|\kappa|$). In this situation, the two split-resonances couple via Floquet mode generation and are no longer independent, as depicted by the black arrows in Fig.~\ref{fig:sideband_contour}a and Fig.~\ref{fig:sideband_contour}b, which show the variation of the first higher-order Floquet mode as a function of the driving frequency and resonance detuning. At lower $\Omega$, each resonance mode (CW and CCW) behaves like an independent resonator, generating Floquet modes that do not lead to inter-modal coupling, and their amplitude decreases as $\Omega$ increases. However, once the generated Floquet mode interacts with the other split resonance, its amplitude increases again due to resonance. Eventually, the Floquet mode will decouple from the split resonance once $\Omega \gg 2|\kappa|$.

The second scenario is when $(|\delta\gamma_b| = |\kappa|, \Delta\phi=\pi)$, $\delta\omega_m \neq \delta\gamma_l$, and $\theta=0$. It represents the cavity operating at an exceptional point. In this case, the $\kappa + \gamma_b=0$ in eqn. \ref{eqn:recursive2} represents the null transfer of optical carrier energy from the $CW$ to the $CCW$ as already discussed in the preceding section. The coefficients of $n\pm1$ are still identical, implying again a symmetric conversion of the Floquet mode, i.e., the red arrows in Fig. \ref{fig:sideband_contour}a will still be bi-directional. However, under EP condition, since the coupling between the split resonances has been made uni-directional (see Fig. \ref{fig:schematic}b), the cavity behaves like an ideal cavity with no resonance splitting, and hence, the Floquet mode conversion shows no signs of resonance splitting based conversion in Fig. \ref{fig:sideband_contour}b. Furthermore, as opposed to the first case, Floquet mode-induced coupling will be restricted in this case due to the forbidden nature of the coupling between CW and CCW modes at the EP, i.e., the black-arrow in Fig. \ref{fig:sideband_contour}a will not exist in this case.

The final case is when $(|\delta\gamma_b| = |\kappa|, \Delta\phi=\pi)$, $\delta\omega_m = \delta\gamma_l$, and $\theta=\pi/2$. Under these conditions, the cavity operates on an EP with an asymmetric Floquet coupling coefficient. While $|S^-|$ is maximized at this point, $S^+$ becomes zero, indicating unidirectional Floquet mode coupling at this specific operational point, as depicted by the leftward red arrows in Fig. \ref{fig:sideband_contour}a for both resonances. Consequently, two unidirectional couplings arise-split resonance coupling due to the EP conditions and Floquet mode coupling due to the modulation parameters. Moreover, as we will demonstrate later, this point corresponds to a null optical carrier transmission, which implies an efficient conversion of optical carrier power to Floquet modes and a doubling of the surviving Floquet mode components. This results in a complete redistribution of the optical input power among the Floquet modes. Ultimately, it effectively translates the carrier frequency to a new frequency component, separated by a spectral distance of $\Omega$.

\begin{equation}
 T^{-1} = \frac{4\delta\omega_m\gamma_c[(\iota\Delta\omega+\gamma)^2-(\iota\Delta\omega+\gamma)(\iota\Omega) + \iota\kappa(\iota\kappa+\iota\gamma_b)]}{[(\iota\Delta\omega+\gamma)^2 - \iota\kappa(\iota\kappa+\iota\gamma_b)][- \iota\kappa(\iota\kappa+\iota\gamma_b) + (\iota\Delta\omega+\gamma-\iota\Omega)^2]}\label{eqn:imp1}
\end{equation}

\begin{figure}[!b]
\centering
 \includegraphics[width=75 mm]{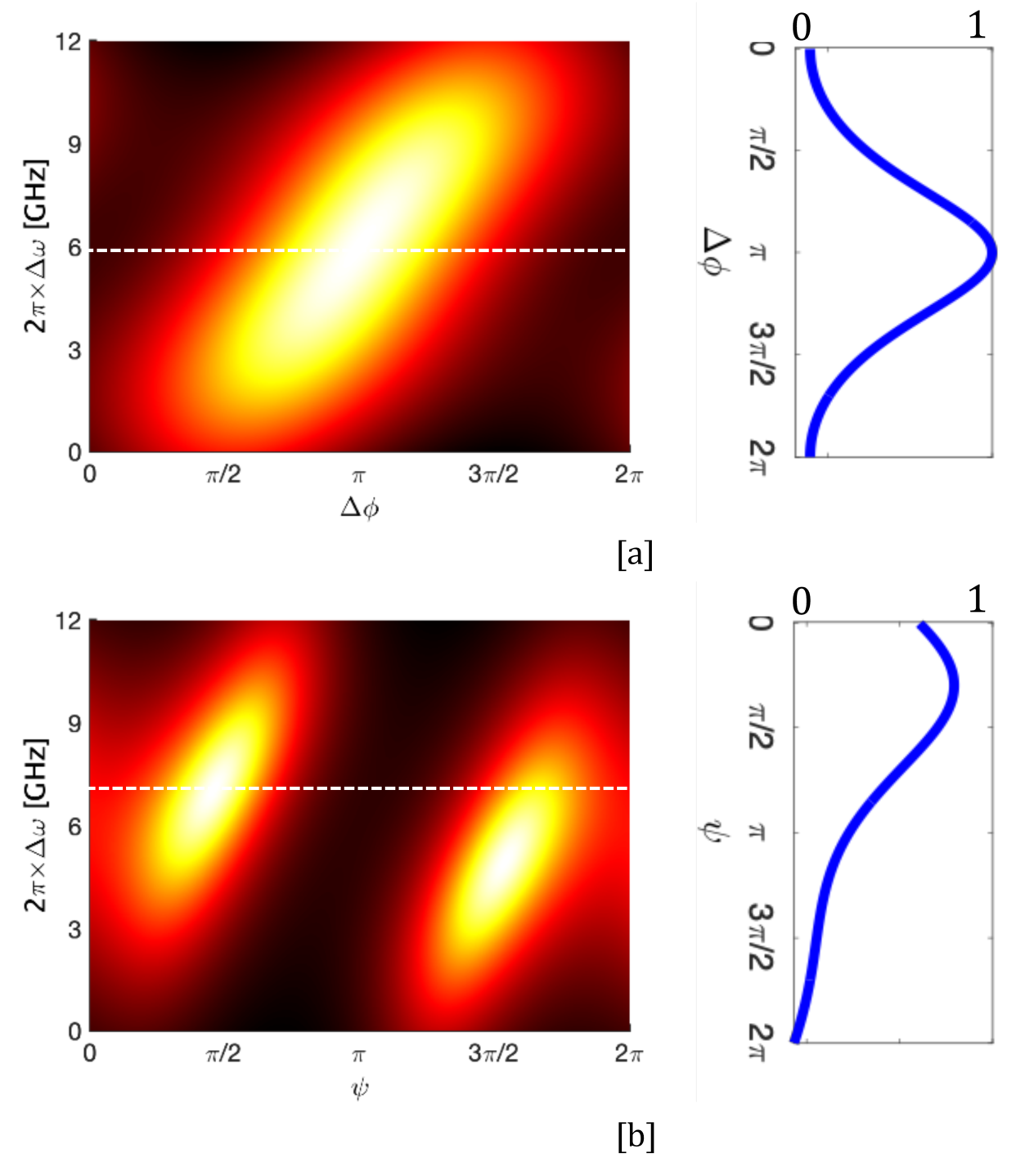}
 \caption{Variation of [a] first lower order Floquet mode conversion on $\Delta\phi$ when $\psi =0$ and [b] $\psi$ when $\Delta\phi = 0$.}\label{fig:imperfection}
\end{figure}

The effect of imperfections in the phase $\Delta\phi$ or $\psi$ on the Floquet mode conversion is done by solving (\ref{eqn:recursive1}) and (\ref{eqn:recursive2}) at $(|\delta\gamma_b| = |\kappa|$, $\delta\omega_m = \delta\gamma_l, \theta = \pi/2$ but varying the values of $\Delta\phi$ and $\psi$. The expression of the transmission of the first lower order FLoquet mode from the CW mode is then given by eqn. \ref{eqn:imp1}. The transmission variation as a function of laser-detuning and $\Delta\phi$ is shown in Fig.~\ref{fig:imperfection}a. The variation as a function of $\psi$ is shown in Fig.~\ref{fig:imperfection}, clearly showing that the transmission is more sensitive to the variation and asymmetric about its maximum value with $\psi$. To ascertain the reason, we obtain the resonance frequency splitting as well as the loss splitting of the cavity resonance as a function of $\psi$ can be approximated as:

\begin{equation}
 Re(\lambda_+ - \lambda_-) = 2|\kappa|\cos(\psi - \frac{\pi}{4})|\sqrt{|\Delta\phi-\psi|}
\end{equation}
and
\begin{equation}
 Im(\lambda_+ - \lambda_-) = 2|\kappa|\sin(\psi - \frac{\pi}{4})|\sqrt{|\Delta\phi-\psi|}
\end{equation}
It can be seen that while the frequency and loss changes depend on the perturbations in the difference between $\Delta\phi$ and $\psi$, the minimum value of the difference exists at different values of $\psi$ for the frequency and the loss. Since the Floquet mode conversion depends on both the source frequency detuning ($\Delta\omega$) and the loss factor (eqn. \ref{eqn:imp1}), the conversion will have a non-trivial effect on both of these factors, explaining the complex dependence pattern on $\psi$. At the EP, the transmission of the carrier, as well as the first-order lower Floquet mode, can be written from eqn. \ref{eqn:imp1} to be:
\begin{equation}
 \alpha_{cw}^{0} = \frac{-\iota\sqrt{2\gamma_c}}{\iota\Delta\omega+\gamma}
\end{equation}
and
\begin{equation}
 \alpha_{cw}^{-1} = \frac{-S^{-1}}{\iota\Delta\omega+\gamma-\iota\Omega}\alpha_{cw}^{0},
\end{equation}
which is the exact expression of $\alpha(t)$ for a cavity if solved under no back-scattering condition and fulfilling the modulation parameters discussed in the third case. It confirms that not only does EP make the cavity performance immune to backscattering but also forces the carrier transmission to be zero at the critical coupling condition ($\gamma_c = \gamma_l$), denoting efficient extraction of the carrier for Floquet mode conversion.

To implement this method experimentally, precise control over the coupling coefficients and phase shifts is required. While achieving exact conditions can be challenging due to fabrication imperfections and material inconsistencies, our approach is tolerant to small deviations, as the EP can still be approached closely enough to observe significant suppression of backscattering. Advanced fabrication techniques and active tuning methods, such as thermo-optic or electro-optic phase shifters, can be employed to fine-tune the system parameters post-fabrication. This post-fabrication flexibility ensures that minor variations in material properties or fabrication tolerances can be corrected. Moreover, the proposed method is fully compatible with existing photonic platforms, as it utilizes well-established components, including ring resonators, Sagnac loops, and phase shifters, all of which are widely adopted in integrated photonics technologies.

Dynamic modulation can be realized using standard electro-optic modulators. For instance, phase modulation can be realized by optimizing the doping profile of the waveguide material, while amplitude modulation can be facilitated through the heterogeneous integration of III-V materials. Both techniques have become standard components in the process development kits (PDKs) of several commercial foundries, ensuring compatibility with established fabrication workflows. The integration of electro-optic modulators within the photonic platform also ensures high-speed performance, making the system suitable for applications in signal processing and optical communication. The combination of static and dynamic tunability ensures that our approach can address a wide range of operational scenarios, from stable, long-term applications to environments requiring rapid, real-time adjustments.

In conclusion, our study demonstrates a viable method for significantly mitigating backscattering in micro-ring cavities by operating at an exceptional point. By harnessing the non-commutative, direction-dependent frequency shifts induced by the chiral transmission properties of non-conservatively coupled degenerate modes, we offer an innovative solution to address backscattering in microcavities. Our detailed theoretical analysis and numerical simulations confirm the efficacy of this approach in suppressing backscattering and enhancing the cavity's function as a modulator. We have also analyzed the impact of parameter variations and imperfections, showing that the method is robust against practical implementation challenges. The versatility and compatibility of our technique with existing integrated photonics platforms equipped with suitable electro-optic modulators highlight its broad applicability. Furthermore, this work paves the way for advancements in diverse fields, such as quantum communication and information processing, and suggests promising avenues for future research and development.

\medskip
\textbf{Disclosures} The author declares no conflict of interest.

\textbf{DATA AVAILABILITY}
The data that support the findings of this study are available
from the corresponding author upon reasonable request.

\bibliography{aipsamp}

\end{document}